# On Laplace-Runge-Lenz Vector as Symmetry Breaking order parameter in Kepler Orbit and Goldstone Boson

Manouchehr Amiri
manoamiri@gmail.com
Health Organization, Tehran, Iran

We introduce a type of symmetry breaking and associated order parameter in connection with Laplace-Runge-Lenz vector of Kepler orbit through an extended spatial dimension and "Ensemble view". By implementation of a small extra spatial dimension and embedded infinitesimal toral manifold, it has been shown that emerging of $LRL$ vector under $SO(4)$ symmetry is in analogy with a variety of explicit and spontaneous symmetry breaking situations and related Goldstone bosons such as phonons and spin waves. A theorem introduced to generalize this concept of breaking symmetry. The diffeomorphism of circular orbit (geodesic) to elliptic one proved to be equivalent with a covariant derivative and related parallel displacement in this extended four dimensional spatial space. Respect to ensemble definition This diffeomorphism breaks the $O(2)$ symmetry of initial orbit and Hamiltonian to $Z_2$ resulting in broken generators in quotient space and associated Goldstone boson as perturbing Hamiltonian term leading to a perpetual circular motion on 2-torus comparable to the perpetual motion idea in "Time Crystal" of Wilczek et.al. This leads to an introduction of gravitational gauge potential under the symmetry of Cartan sub algebra of $su(2)$.

Introduction

Reasonable evidences in Hamilton–Jacobi theory states that many mechanical problems could be reduced to the force free motion on non-Euclidean manifolds, and one may model a system with negative energy (bounded orbit) with force free motion on a certain hypersurface in phase space [1,2,**37**,34,35]. On the other hand, general relativity demonstrates the gravitational force as a curvature of space-time without direct distant action of the field source, thus the force exerting on an orbiting particle in a central gravity field, comes from the geometric structure of space-time and is comparable to tidal force of free motion on non-Euclidean (Riemannian) manifold which corresponds the motion of small particles on geodesics [3, 4].
we apply and generalize these ideas to show that motion on a hypersurface with an extra spatial dimension under a group symmetry interprets the emerging $LRL$ vector in elliptic orbit as a result of a rotation (or boost) of a generalized angular momentum in such a higher dimension and conclude that the projection of the initial circular orbit under this rotation group gives rise to a symmetry breaking and elliptic orbit. Goldstone Boson as a result of local or global broken symmetry manifests a spectrum of scalar or vector bosons in a various setting of breaking symmetry with degenerate ground state. We generalize this implication to a large scale as is seen in Kepler orbit. Using an 'Ensemble view' and under some boundary conditions, considering a small extra spatial dimension and $so(4)$ as governing algebra of spatial group symmetry, we derive symmetry breaking of Kepler orbit and show that the boost like rotation along extra spatial dimension results in breaking symmetry in conventional spatial 3-dimension. Transforming the circular orbit to an elliptic one, modelled by orbit diffeomorphism to geodesic on infinitesimal 2-torus, proved to be equivalent with a parallel displacement and involves contorsion as a degree of freedom in extra dimension. Then the broken generators and associated Goldstone boson will be introduced as perturbing Hamiltonian term leading to a perpetual circular motion on 2-torus. Consequently $LRL$ vector emerges due to invariance of total Hamiltonian under $so(4)$ algebra symmetry. A modified algebra and related root space connects the charges (chemical potential) with the field of Goldstone boson.
This suggests that presumption of a compacted small extra spatial dimension may be reasonable to introduce new era in large scale physics in order to exploit the concept of symmetry breaking. As a cornerstone, symmetry breaking in various field of modern physics acting an influential role in understanding of many implications as particle mass, pions, magnons (spin wave) and phonons [5,6]. This symmetry breaking occurs in internal or external symmetries corresponding to internal or external coordinates. Implication of Goldstone bosons that arises as consequence of breaking of continuous symmetry was born in the high energy physics and standard model and found interesting applications in various fields of physics such as particle and condensed matter, phase transition and solid state [7, 8]. Examples of internal Symmetry breaking include *chiral* symmetry breaking of strong interaction with resulting (Nambu) Goldstone Bosons realized as *pions*. In standard model $SU(2) \times SU(1)$ gauge symmetry breaking with Higgs mechanism generates masses to standard model related particles [9], and $U(1)$ gauge symmetry breaking became a theoretical model in superconductivity. *Magnons* also represent the Goldstone bosons of the spontaneously broken symmetry of rotation group $O(3) \to O(2)$ in (anti) ferromagnetism. The latter example denotes a global symmetry breaking vs. the other examples of local symmetries. In this article we investigate a global symmetry breaking pertained to Kepler problem exploiting an "Ensemble view" method which has not been identified before. Certainly this approach can be applied for Kepler orbit because of its *periodic* nature. As usual we use the standard definition of *time ensemble* for probability distribution $p(\vec{r})$ of particle position on orbit. Therefore one may imagine an infinite sequences of copies of this system at different time [10] (cf. Boltzmann 1894). This "Ensemble view" could be predicted for any periodic phenomena. For convenience we prepare this ensemble by determining the particle position in an infinitesimal time interval $\Delta\tau$ for each copy. In such an ensemble, after overlapping all copies in ensemble, we find a definite position probability for orbiting body. Obviously in a circular orbit the related probability distribution will found to be a uniform one with $O(2)$ symmetry; however in an ellipse orbit this distribution evolves to a $Z_2$ symmetry respect to the major axis of ellipse. Obviously this ensemble is equivalent to the ensemble of infinite copies of identical Kepler problem recorded at an infinitesimal interval $\Delta\tau$. Symmetry breaking in this setting has the strict similarity with symmetry breaking in (anti) ferromagnetism because in (anti) ferromagnetism we can use equivalently the time ensemble view instead of space ensemble and consequently instead of a large number of atoms in external magnetic field we will use *time-ensemble* of a single atom with a probability obtained by Boltzmann distribution proportional to $e^{\frac{-\mu B}{kT}}$. Interestingly the Lagrangian symmetries of these mechanical settings are identical with probability distribution symmetries in ensemble. We study the symmetry breaking of circular orbit to an ellipse with infinitesimal eccentricity under energy conservation leading to generation of a new motion integral, Laplace-Runge- Lenz vector. Historically this vector analyzed in detail by Fock and Pauli in classical and quantum problems i.e. Kepler orbit and Hydrogen atom by $SO(4)$ symmetry [11, 12].

## Section 1

### 1.1.1 Symmetry Breaking in Kepler orbit

Suppose an exact circular orbit of Kepler problem as has been shown in some solutions of Schwarzschild equation with spherical symmetric gravitational field. For the sake of simplicity the field source and orbiting planet both be considered point like masses. In this setting and under energy conservation, we consider an infinitesimal change implemented on the orbit toward an elliptic one with an infinitesimal eccentricity. Let assume a time ensemble of this Kepler orbit (copies of Kepler system) in which we consider an ensemble of huge number of Kepler system copies at different time intervals $\Delta\tau$ with a central field and a rotating point like mass orbiting bounded in the central field. Probability density to find the point mass in a random infinitesimal interval of time $\Delta\tau$ at each point on the orbit in this *time-ensemble* is proportional to inverse of velocity at that point:

$$p(\vec{r}) = \frac{1}{V(\vec{r})} \qquad (1.1)$$

Whenever the orbit has the full circular symmetry, the microcanonical ensemble probability to find the mass in an infinitesimal time interval $\Delta\tau$ is identical at all points of the orbit as a result of constant velocity. However a small perturbation toward ellipse orbit while Hamiltonian remains invariant leads breaking uniform probability distribution to a 2outcome distribution because the probability to find the mass in each semi orbit on the opposite side of long ellipse axis is equal to $\frac{1}{2}$ and probability density on each point of the orbit is the same as the mirror image point respect to the long axis of ellipse. So the group symmetry of probability distribution of ensemble and Lagrangian breaks from $G = O(2)$ in circular orbit to subgroup $\tilde{G} = Z_2$. This evolution as diffeomorphism from circular to elliptic orbit can be viewed as a symmetry breaking. In circular orbit the symmetric group of Lagrangian $\mathcal{L}$ (or Hamiltonian $\mathcal{H}$) denoted as rotation group $G = O(2)$ while in elliptic orbit this symmetry reduces to a $Z_2$ symmetry i.e. the invariance under mirror symmetry respect to the long axis of ellipse. As a conclusion, symmetric group of Lagrangian $\mathcal{L}$ and ensemble states probability distribution in circular and elliptic orbit is the same. We will exploit this property later in this article. The quotient group $G/Z_2$ characterizes the broken generators and the spanning space of Goldstone bosons. More analyzing 'Ensemble view' will be discussed in sec (1.1.3).

### 1.1.2 Lorentz transformation and symmetry breaking

It is possible having a thought experiment of a closed system of free particles and their rest frames as *parallel* reference frames set up in stationary state i.e. all particles and their rest reference frames are relatively motionless. In this setting there exists a full symmetry in transition from a reference frame to another one. Now if we impose a motion with small velocity $v$ along some axis to one of these reference frames $\mathcal{R}$ (or particle) while preserving the parallelism between reference frames, this symmetry will be broken because the moving reference frame actually is no longer parallel relative to the other reference frames. The reason is Lorentz transformation interpretation and associated boost rotation of $\mathcal{R}$ in $x_0-x_i$ plane with $x_0$ as time and $x_i$ as one of spatial axes. In other words a small linear velocity results in a relative boost rotation of moving frame $\mathcal{R}$ respect to all other frames and therefore in this system, there is a preferred direction along the motion trajectory. This breaking symmetry in some sense resembles the spatial symmetry breaking in crystals where the ordering atoms in some direction break the whole symmetry, or symmetry breaking in ferromagnet because of ordering atoms along the external magnetic field $\vec{B}$.

Moreover $\mathcal{R}$ observers detect the circles on other reference frame as an *ellipse* due to the so called Lorentz transformation which means the breaking symmetry of $O(2)$ to $Z_2$ and this change is also a consequence of the aforementioned $x_0-x_i$ boost rotation. The eccentricity of this ellipse can be calculated to be equal to $v/c$. This picture could also be well understood when an electric charge with a spherical symmetric electric field located on $\mathcal{R}$. In such situation the breaking symmetry after motion of $\mathcal{R}$ results in a new field (or charge) that obeys the cylindrical (axial) symmetry i.e. magnetic field $\vec{B}$. In summary a spatial symmetry breaking is equivalent to a boost like rotation on $x_0-x_i$ plane as *extra dimension* that preserves the new metric of 3+1 dimension as an isometric transformation. This is the basic idea to develop an extra dimension in presenting model. This will be presented as a theorem in (sec1.7). We use a general formalism in which a *conserved charge* always associated with symmetry breaking. This charge in Lorentz transformation and its associated symmetry breaking is simply the conserved velocity vector of $\mathcal{R}$, in Ferromagnets their magnetic moment and in Kepler problem *LRL* (Laplace-Runge-Lenz) vector.

### 1.2 Emerging LRL Vector

Transition of orbit from circular to ellipse, also results in emerging *LRL* vector (Laplace-Runge-Lens), an integral of motion that as a conserved charge cannot be attributed to an explicit Lie algebra and group symmetry [13, 14].on the other hand searching for the symmetry governing the invariance of *LRL* often reveals a type of internal symmetry [15] and in some articles reminded as hidden, accidental or dynamical symmetry [2, 16]. After Fock and Pauli [1, 30, 31] $SO(4)$ accepted as the group of hidden symmetry of *LRL* vector in Hydrogen atom and Kepler problem while the orbiting planet is bounded (i.e. $E < 0$) [1, 2]. Some of other simple dynamical systems with hidden symmetry can be found in the works of Aronson et.al [32] and Ghirardi [33].There is still controversy about the exact symmetry in *LRL* problem. $SO(4)$ as a symmetric group acts on 4 dimensional space through 2 set of generators $J_i$ and $K_j$ ($i$, $j = 1,2,3$).

$$
\begin{aligned}
&[J_p, J_q] = i\varepsilon_{pqr} J_r \\
&[J_p, K_q] = i\varepsilon_{pqr} K_r \\
&[K_p, K_q] = i\varepsilon_{pqr} J_r
\end{aligned}
\qquad (1.2)
$$

$J_i$ denoted as the angular momentum components and $K_i$ as *LRL* vector (or parallel to it). The generators of $so(4)$ after a linear combination of the basis split to two independent $su(2)$ subalgebras.

$$A_i = \tfrac{1}{2}J_i + \tfrac{1}{2}K_i \qquad B_i = \tfrac{1}{2}J_i - \tfrac{1}{2}K_i$$

Thus its Lie algebra is isomorphic to $su(2)_R \oplus su(2)_L$ with independent basis $A_i$ and $B_i$ respectively.

$$[A_i, A_j] = i\varepsilon_{ijk} A_k \quad , \quad [B_i, B_j] = i\varepsilon_{ijk} B_k \quad , \quad [A_i, B_j] = 0 \quad (1.3)$$

Consider a Kepler problem with exact circular orbit without *LRL* vector. It is well known that the *LRL* vector $A$ is proportional to the eccentric constant of ellipsoid orbit and as far as the orbit is circular, *LRL* vanishes. This could be verified by equation [2]:

$$|A| = mke$$

With $e$ as eccentricity and $m$ as mass. Now imagine an infinitesimal change of circular orbit of an orbiting point like mass $\mu$ toward ellipse orbit with conserved energy. This evolution accompanies an infinitesimal *LRL* vector and along invariance of energy we may write down the relation of *LRL* vector with angular momentum and energy as follows [2]:

$$|A| = mk\sqrt{1 - \frac{2L^2 E}{mk^2}}$$

With $L$ and $E$ as angular momentum and energy of mass. For the sake of simplicity we set $m = 1$ and $E = \frac{1}{2}$ then we have:

$$A^2 + L^2 = k^2 \quad (1.3')$$

Therefore in Hamiltonian invariant transformation $A^2 + L^2$ will be conserved.

### 1.3 Interpretation of $A_i$ and $B_i$

Although there exists a well-defined interpretation for $J_i$ and especially $J_3$ as the generator of rotation, there are no explicit physically substitution for $A_i, B_i$ and bases [2, 9]. Hence the symmetry pertained to conservation of *LRL* vector remains unclear and refers as dynamical or hidden symmetries [11]. *We will show in the next sections that this linear combinations of so(4) i.e. $A_i, B_i$ can be interpreted as two su(2) or so(3) as symmetries in a hyperspace with one extra spatial dimension on an infinitesimal torus in such a way that $A_3, B_3$ act on the $R^{3+1}$ manifold (3 usual and 1 extra dimension $x_4$) presents the rotation respect to a new axis resulting from the infinitesimal rotation of axis $x_3$ in the hyper plane $x_3 - x_4$ and consequently this guarantees the Hamiltonian invariance and emerging of a new integral of motion, LRL vector.* Thus *Existence of a spatial extra dimension* with small range similar to Kaluza-Klein, Ads- De sitter space or Einstein-Bergmann 5D model [21] could resolve this controversy definitely.

### 1.4 Extra spatial Dimension

For the sake of simplicity we choose 2 dimensional spatial subspace, $x_1 - x_2$ plane as the plane of orbiting in Kepler problem. In this setting $J_3$ acts as generator of rotation in the plane around the $x_3$ local axis. We add an extra dimension with small range $\mathfrak{r}$ to the usual $M^4$ space-time. This space can be regarded as $\mathfrak{S} = M^4 \times \mathfrak{r}$ and allows the action of $SO(4)$ group on this manifold. Because of the small range of the extra dimension, one may assume a *flat dimension* for its structure. As can be seen in $SO(3,1)$ Lorentz group, $K_i$ may be viewed as Boost generator with a difference that in this case, rotations are in the plane of extra dimension axis and one of the usual spatial axes Fig (1).

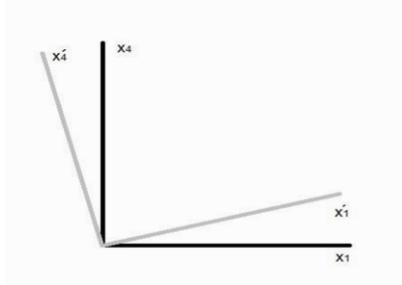

Fig (1): Boost like rotation in $x_1$- $x_4$ plane.

To visualize the situation we substitute the extra spatial dimension $x_4$ instead of the $x_3$ local axis orthogonal to the $x_1$-$x_2$ plane with the origin of coordinate located at central field source and exploit the time as a parameter. $K_3$ generates boost like rotations in the $x_3$-$x_4$ plane and under such a rotation, the projection of the rotated circular orbit onto the $x_1$-$x_2$ plane will become an ellipse with a small eccentricity. Moreover the circular configuration will preserve in 3+1 spatial dimension. If a Kepler circular orbit undergoes an infinitesimal diffeomorphism while leaves the energy constant resulting in an elliptic orbit with small eccentricity $\varepsilon$, then a small *LRL* vector $\delta A$ appears orthogonal to new angular momentum $\vec{L'} = \vec{L} - \delta \vec{A}$. Respect to (1.3') we obtain:

$$|\vec{K}|^2 = |\vec{L}|^2 = |\vec{L'}|^2 + |\delta\vec{A}|^2 \qquad (1.4)$$

Where $\vec{K}$ presented as a 4-vector which lives in 3+1 spatial dimension and we label it as "generalized angular momentum" while for circular orbit we have $\vec{K} = \vec{L}$ and after infinitesimal evolution reads as:

$$\vec{K} = (L'_1, L'_2, L'_3, \delta A) \in \mathfrak{S} \qquad (1.5)$$

All four components are spatial, 3 in regular space and one along extra dimension. So $\delta A$ refers to the component along the extra dimension. The evolution from circular to elliptic orbit is the result of the effect of $so(4)$ generators with very small parameters on $\mathfrak{S}$:

$$\vec{K'} = (1 + \varepsilon_i \hat{K}^i)\vec{K} = \vec{L'} + \delta\vec{A} \qquad (1.6)$$

So the circular orbit and related generalized angular momentum $\vec{K}$ undergo a very small rotation through the extra dimension while the $\vec{K}$ module (and energy) remains unchanged and regular angular momentum $\vec{L'}$ appears as the projection of $\vec{K}$ onto the 3d spatial space ($x_3$ axis). Because of the limited range of the extra-dimension, successive action of $(1 + \varepsilon_i \hat{K}^i)$ in the usual way is impossible and after a small rotation inside the range of extra dimension it cannot be continued. However the evolution will be carried out by action of this operator on the projection of rotated orbit onto the orbiting plane and this cycle repeats up to a definite value of rotation. We call this transformation as *sequential infinitesimal rotation* in $\mathfrak{S}$. This simple interpretation guarantees the energy conservation and introduces the existence of *LRL* vector. On the other hand by using Kepler 3rd law the angular momentum conservation remains valid for $\vec{L'}$.
In the following sections this transformation realized as the motion on the higher dimensional manifolds.

1.5.1 Free Particle motion on the manifold

As mentioned earlier, most of mechanical problems can be reduced to the force free motion on non-Euclidean manifolds [37]. This motion on manifold's geodesic could be attributed to a particle, center of mass of many particles system or any dynamical system with relevant conditions.
The previous transformation as *sequential infinitesimal rotation* in spatial sub space of $\mathfrak{S}$ is a straight forward method to geometric description of *LRL* emergence. In this section we show the equivalence of this transformation with an *infinitesimal toric diffeomorphism* $\delta\mathfrak{T}$ and a specific type of *"Covariant derivative"*.
Consider circular Kepler orbit on the *xy* plane as the axis of a 2-Torus lying on the plane and an orbit with winding number 1:1 on this torus (Fig2). Then the following statements could be taken into account:

*Corollary1:*

*Considering a 2-torus and a 1:1 orbit (winding number =1) on it, If the small radius of torus approaches zero, $\varepsilon \to 0$ then at the limit converges to the geodesic on the torus with the same winding number and , at $\varepsilon = 0$ the 1:1 orbit coincides circular axis. Inversely a circular orbit as 2-torus axis evolves to 1:1 orbit (or geodesic) when $\varepsilon$ grows from zero to a definite small value* (Fig2).

This corollary is obvious because at the limit $\varepsilon \to 0$ the metric on this torus approaches the flat metric like the flat 2-torus. Then conversely we could imagine a diffeomrphism of circular orbit toward a 1:1 orbit or geodesic on torus with a very small radius $\varepsilon$ which in turn converges to the geodesic with winding number 1:1. We call this transformation as $\delta\mathfrak{T}$ and show the projection of this orbit onto the *xy* plane is an ellipse with eccentricity $\varepsilon$.

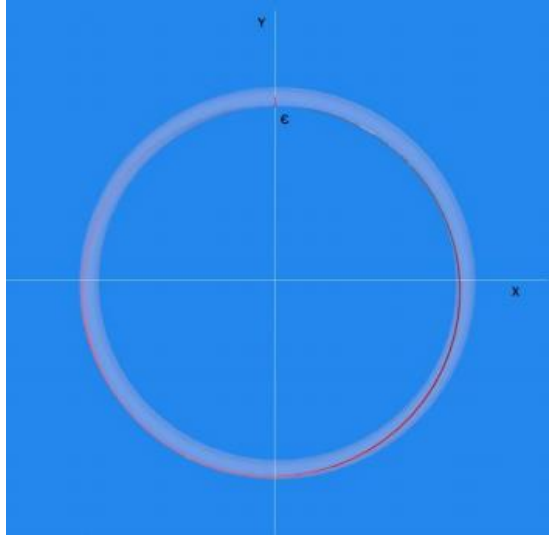

Fig (2): projection of orbit with 1:1 winding number of 2-torus onto **xy**-plane presented as an ellipse with long axis along **y** and eccentricity $\varepsilon$. The range of extra dimension assumed to be about $\varepsilon$ in the orthogonal direction to the **xy**-plane. The orbit can be obtained from circular axis of torus by displacement $\varepsilon$ (along negative **y** axis) and rotation $\delta\theta = \frac{\varepsilon}{R}$ through the extra dimension.

In our model transformation of circular to ellipse orbit (with small eccentricity) resembles the length contraction in Lorentz transformation that leads the circular geometric shapes transforms to ellipses.

*Corollary2:*

*Projection of the infinitesimal 1:1 orbit (geodesic) of 2-Torus (embedded in **x** , **y** and extra dimension) onto **xy** plane is an ellipse with eccentricity equal to small radius $\varepsilon$ . This means that after infinitesimal transformation of circular orbit to geodesic on 2-Torus, the projection onto the usual spatial dimension will become the diffeomorphism of initial orbit with invariant Hamiltonian* Fig (2).

Due to Clairaut theorem the angular momentum of free particle orbiting about a central field on an axial symmetric manifold will be conserved. One concludes that free motion of particle on 1:1 geodesic lying on 2-Torus leaves the *generalized angular momentum* conserved in whole space $\mathfrak{S}$. Therefore the projection of the equal time sectors of 2-torus orbit in $\mathfrak{S}$ onto *xy* plane manifests equal area due to the simple projected area calculation. Therefore the angular momentum on *xy* plane $\vec{L'}$ will remain conserved and determined by equation (1.4).

This completes the model, the transformation $\delta\mathfrak{T}$ results in a new projected orbit on *xy* plane while the Hamiltonian and angular momentum remain conserved.

In this limit interestingly we found that this evolution of circle to geodesic on infinitesimal 2-torus is the same as the action of $(1 + \delta\theta_i \widehat{R}^i)$ plus $\varepsilon_i \partial^i$ with equal parameter $\varepsilon_i = R\delta\theta_i$ while $\varepsilon_i$ still remains very small $\varepsilon_i \ll 1$ and stated as eccentricity of elliptical orbit. Then:

$$\delta\mathfrak{T} = [\widehat{\mathbf{1}} + \varepsilon_i R^{-1}\widehat{R}^i + \varepsilon_j \partial^j] \qquad (1.7)$$

Where *R* is the radius of circular orbit and is too large compared to the order of extra dimension $\mathfrak{r}\sim\varepsilon$. *This equation implies the equivalence of a Poincare like transformation and geodesic diffeomorphism of the initial orbit.*

This equivalence also conveys us that the $\delta\mathfrak{T}$ transformation may be viewed as a *Parallel transport.*

Combination of a translation and rotation (or boost) is the fundamental transformation in quantum field theory, the so called Poincare transformation, but here the rotation component takes place along an extra dimension.

Take into account an infinitesimal transformation of rotation across an axis with an angel $\varepsilon R^{-1}$ plus displacement along the same axis by $\varepsilon$:

$$[\hat{1} + \varepsilon R^{-1}\hat{K}^i + \varepsilon \partial^i] = \hat{1} + \varepsilon(R^{-1}\hat{K}^i + \partial^i) \quad (1.8)$$

We assess the closure property of these elements for $\varepsilon, \mu \ll 1$:

$$[\hat{1} + \varepsilon(R^{-1}\hat{K}^i + \partial^i)][\hat{1} + \mu(R^{-1}\hat{K}^i + \partial^i)] = \hat{1} + (\varepsilon + \mu)(R^{-1}\hat{K}^i + \partial^i) \quad \text{(Closure property)}$$

Terms containing $\varepsilon\mu$ and $R^{-2}\varepsilon\mu$ are of order $o(2)$ and have been vanished. Other group properties are also could be verified. Now it seems that the generator $(\partial^i + R^{-1}\hat{K}^i)$ after lowering indices by contraction:

$$\hat{D}_j = g_{ij}(\partial^i + R^{-1}\hat{K}^i) = (\partial_j + R^{-1}\hat{K}_j)$$

Plays the "Parallel Transport "or "Covariant Derivative" role equivalently. In the context of Einstein-Cartan space one could observe that the motion is equivalent to a "rotation" and "displacement" [22]. If the displacement happens along $x^1$ and rotation in $x_1-x_4$ plane under the effect of $\hat{K}^3$, $\Gamma_{k1}^j$ serves as the connection on the 2-torus and then:

$$\delta x^1 = \varepsilon$$

$$g_{ij}\Gamma_{k1}^j v^k dx^1 = \lambda_\mu^{3\nu} g_{\nu k} v^k R^{-1} dx^1 \Rightarrow R g_{\mu j}\Gamma_{k1}^j = g_{\nu k}\lambda_\mu^{3\nu}$$

$$R\Gamma_{\mu k 1} = g_{\nu k}\lambda_\mu^{3\nu}$$

Product of the last equation by $g^{k\beta}$ gives:

$$R\Gamma_{\mu 1}^\beta = \delta_\nu^\beta \lambda_\mu^{3\nu} = \lambda_\mu^{3\beta} = (\lambda^3)_\mu^\beta \quad (1.9)$$

$(\lambda^3)_\mu^\beta$ is the matrix representation of $\hat{K}^3$. Recalling the isospin connection with $(\tau^a)_{ij}$ as Pauli matrices and $A_\mu^a(x)$ as gauge potentials which reads as [36]:

$$\Gamma_{ij1} = -\frac{i}{2}A_1^a(x)(\tau^a)_{ij} \quad (1.10)$$

And product of both sides by $g^{jk}$ and assigning $A_1^3 = -\frac{1}{R}$ and $A_1^1 = A_1^2 = 0$, results in:

$$R\Gamma_{i1}^k = -\frac{i}{2}(\tau^3)_i^k \quad (1.11)$$

Comparison of equations (1.9) and (1.11) reveals their strict similarity and gauge potential $A_1^3(x)$ can be Interpreted as the usual classic Gravitational scalar potential i.e. $A_1^3 = -\frac{1}{R} \approx \varphi$

This relation reveals a common formalism in quantum and celestial level and interpretation of graviation as a gauge potential. Although $su(2)_R$, $su(2)_L$ bases generally are *anholonomic*, we accept $\hat{K}_3$ and $\hat{J}_3$ as *holonomic bases* or *Killing vectors* on tangent spaces of infinitesimal torus because as we showed earlier $\hat{K}_3 = \partial_\theta$ and $\hat{J}_3 = \partial_\varphi$ are Cartan subalgebra and commutative:

$$[\partial_\theta, \partial_\varphi] = 0 \quad (1.12)$$

Because $R$ is large compared to $\varepsilon$, curvature of gravity is so small and Christoffel symbol i.e. the affine connection $\Gamma_{\mu 1}^\beta$ values are very small and slow varying quantities therefore $\Gamma_{\mu j}^\beta \Gamma_{\nu k}^\alpha$ and $\Gamma_{\nu\sigma,\rho}^\lambda$ are negligible of order $O(2)$: $\quad \Gamma_{\mu j}^\beta \Gamma_{\nu k}^\alpha \approx 0 \quad , \quad \Gamma_{\nu\sigma,\rho}^\lambda \approx 0 \quad (1.13)$

This result is a consequence of the fact that $g_{\nu k}$ values are approximately slow varying function on torus range, then $\Gamma_{\mu j,r}^\beta$ and $\Gamma_{\mu j}^\beta \Gamma_{\nu k}^\alpha$ values approaches to $o(2)$ order and also negligible, then by definition:

$$R^\lambda_{\nu\rho\sigma} = \Gamma^\lambda_{\nu\sigma,\rho} - \Gamma^\lambda_{\nu\rho,\sigma} + \Gamma^\alpha_{\nu\sigma}\Gamma^\lambda_{\alpha\rho} - \Gamma^\alpha_{\nu\rho}\Gamma^\lambda_{\alpha\sigma} \approx 0 \qquad (1.14)$$

$R^\lambda_{\nu\rho\sigma}$ is a negligible value. Recalling the commutation relation of covariant derivatives with torsion:

$$[\nabla_\mu, \nabla_\nu]V^\rho = R^\rho_{\sigma\mu\nu}V^\sigma - T^\lambda_{\mu\nu}\nabla_\lambda V^\rho \qquad (1.15)$$

After vanishing the first term on the right side of equation we obtain:

$$[\nabla_\mu, \nabla_\nu]V^\rho = -T^\lambda_{\mu\nu}\nabla_\lambda V^\rho \qquad (1.16)$$

And in operator form we have: $\quad [\nabla_\mu, \nabla_\nu] = -T^\lambda_{\mu\nu}\nabla_\lambda$

This resembles the commutation relation of Lie algebra with generators $\nabla_\mu$ and structure constants $-T^\lambda_{\mu\nu}$.

On the other hand we know that for a group manifold of $G$ the main connection reads as:

$$\omega^k_{ij} = \bar{\Gamma}^k_{ij} = \frac{1}{2}(c^k_{ij} - c^i_{jk} + c^j_{ki}) \qquad (1.17)$$

With the definition:

$$T^k_{ij} = \Gamma^k_{ij} - \Gamma^k_{ji}$$

After substitution in above equation we obtain:

$$T^k_{ij} = c^k_{ij} \qquad (1.18)$$

This means the torsion tensor is the same thing as structure constant of the Lie algebra whose group manifold is 2 torus of the model that realizes the subgroup with generators $\widehat{K}_3$ and $\widehat{J}_3$. It should also be reminded that structure constant $c^k_{ij}$ behaves as tensor when the algebra bases are transformed under a transformation. This results in the conclusion that $\nabla_\mu$ covariant derivatives are the general form of the diffeomorphism group of transformation of geodesic during breaking the circular orbit symmetry and its constant probability distribution (respect to ensemble probability) toward elliptic orbit with axial ($Z_2$) symmetry of probability distribution.

### 1.5.2 Ensemble view

There are many examples of symmetry breaking in condensed matter such as superconductivity, superfluidity and ferromagnetism that exploiting the ensemble view. In these situations degenerate states form an "ensemble" of states and these states may be attributed by some unitary transformation. In this sense, Goldstone boson presented as collective states or composite particles.

After symmetry breaking, the non-degenerate ground state (i.e. vacuum expectation value) shifts to degenerate states which form an 'Ensemble' of states. We define a similar ensemble for Kepler orbit. First consider circular orbit. In this setting for the observer attaching to $C.O.M$ we have:

$$p_\mu = \partial_\mu = 0 \ (\mu = 1,2,3) \text{ and } p_0 = \partial_0 \neq 0$$

After a perturbation while the Hamiltonian remains invariant, as explained in sec 1.5 and respect to the equation (1.6) observer detects a rotational motion with generator $\widehat{K}^3$. Because of periodic nature of Kepler orbit, and 1:1 winding of new orbit on 2-torus, after one revision of orbit, exactly one revision with small radius $\varepsilon$ takes place. Hence getting ensemble for Kepler orbit is equivalent to ensemble obtained from all states of orbiting particle on small orbit with radius $\varepsilon$, so the primary ensemble discussed in sec 1.1.1 is the similar ensemble for degenerate ground state as seen in symmetry breaking in aforementioned examples i.e. superconductivity,etc.

Therefore in "Ensemble view" we apply the symmetric group on an ensemble of a periodic process such as Kepler orbit and the symmetry breaking also occur in this setting by reducing of the main group to one of its subgroup.

### 1.6 Symmetry breaking

The motion of orbiting mass or center of mass along the circular orbit is geodesic motion which minimizes the action. In the view of an observer in the reference frame attached to $C.O.M$, the orbiting mass occupies the *ground state*. This ground state isn't a degenerate one Fig (3a).

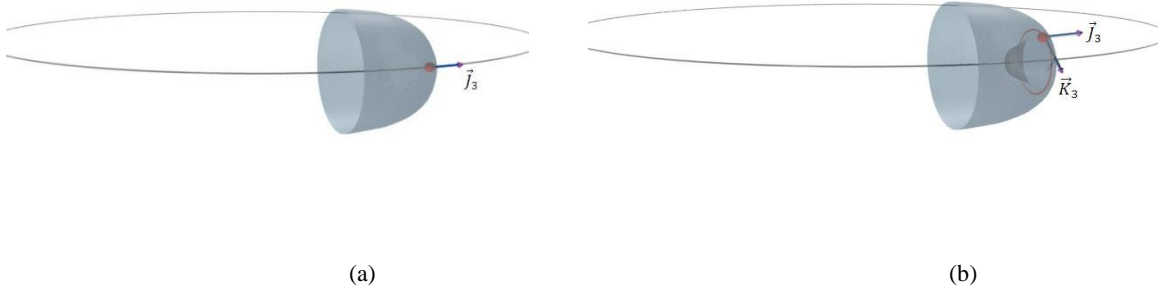

(a)                                      (b)

Fig (3): a) Motion on circular orbit with non-degenerate potential .b) Motion on the degenerate potential states depicted as red circle. Orbiting mass follows a spiral trajectory while bounded to Mexican cap potential well with a winding number 1:1. The trajectory in fig (2) represents the mass orbit in $\mathfrak{S}$. $\widehat{K}_3$ and $\hat{J}_3$ denoted as Killing vectors (or generators) tangent to 2-torus as group manifold. Dark line lies on the 3 dimensional spatial space, while torus embedded in $\mathfrak{S}$ . Projection of this spiral curve onto the 3 dimensional space would be an elliptic orbit as depicted in Fig (2).

After small perturbation in the orbit trajectory, due to our model, the primary symmetry $SO(2)$ in spatial subspace of $M^4$ breaks to $Z_2$ by diffeomorphism of circular orbit (geodesic on $M^4$) to *1:1* geodesic on the torus in $\mathfrak{S}$, Fig(3b). Actually symmetry breaking takes place in $M^4$ rather in $\mathfrak{S}$. The space $\mathfrak{S}$ just undergoes a boostlike rotation with $\widehat{K}_3$ , leaving the Hamiltonian and angular momentum invariant and results in a new invariant *LRL* vector.

An observer on world line in $M^4$ (i.e. the axis of torus) with reference frame attached to $C.O.M$ in *Ground state*, after $\delta\mathfrak{T}$ transformation, encounters a new rotational motion of this mass through extra dimension space, and therefore the non-degenerate ground state transform to a degenerate one. All the points on this circle correspond to minimal energy or ground states; this motion is just the motion on the ground states as like as the motion on Mexican hat hollow in $su(2) \times u(1)$ symmetry breaking of standard model and conveys that the condition for symmetry breaking satisfies and we can exploit this symmetry breaking in Kepler problem. Suppose the coordinate $x^1$ and $x^4$ on $x^1 - x^4$ plane orthogonal to the world line on $M^4$ at the time origin $t = 0$ Fig (3b). Strict similar setting can be observed in symmetry breaking in ferromagnetism and related symmetry breaking $O(3) \rightarrow O(2)$ 'where the vacuum (ground) state corresponds to the state with spin in the direction of the magnetic field, then the *magnons* correspond to excitations caused by flipping of spins [7]. In our model vacuum state is in circular orbit Fig (3a) and excited states the spiral orbit on 2-torus after $\delta\mathfrak{T}$ perturbation Fig (3b).

In view of this observer , there exists a perpetual motion on a circle and oscillation with a frequency or angular velocity $\omega_1$ equal to the mean angular velocity of the orbiting particle in Kepler orbit, resulting in a 1:1 orbit (geodesic) Fig(3b). Inspiring recent papers about "Time Crystal" of Wilczek [29] focusing on the time translation symmetry breaking, the perpetual motion around a circular pattern at ground state is possible. We suggest the similar "rotation" on the ground state in $x^1 - x^4$ plane Fig (3b).

To make this we consider the local field $\varphi = x^1 + ix^4$ to identify the ground state by the condition. This choice is more understandable when we recall the field decomposition in Higgs mechanism symmetry breaking $\varphi = \varphi^1 + i\varphi^4$ substituting $\varphi^1, \varphi^4$ by coordinate variables $x^1, x^4$ .by exerting the condition:

$$(x^1)^2 + (x^4)^2 = \varepsilon^2 \qquad (1.19)$$

In polar coordinate we obtain:

$$\varphi = \varepsilon e^{iv}$$

Where $\varepsilon$ is the small radius of torus. Then we can substitute $v$ local scalar field instead of $\varphi$. Related Lagrangian in this situation reads as:

$$\mathcal{L} = -\frac{\varepsilon^2}{2}(\partial^\mu v)(\partial_\mu v) + m^2\varepsilon^2 \qquad (1.20)$$

Which is invariant under $u(1)$ or equivalently $\widehat{K}_3$ . This symmetry leads to an invariant current simply reads as:
$$\mathcal{J}_0 = -\varepsilon^2 \partial_\mu v \qquad (1.21)$$

Restriction to time displacement by $\partial_0 v = \omega_1$ results in:

$$\mathcal{J}_0 = -\varepsilon^2 \omega_1$$

The charge of related generator reads as:

$$Q = \int_V \mathcal{J}_0 dy = 2\pi\varepsilon^2 \omega_1 \qquad (1.22)$$

This presents the regular angular momentum about motion axis. If $\pi\varepsilon^2$ interpreted as circle area, its equivalent in electric current is magnetic moment defined by $\mu = IS$. Taking into account the light velocity as the upper limit, we conclude a maximum angular velocity obtained by:

$$\omega_{max} = \frac{C}{|v|}$$

Where $C$ is light velocity. *We identify $v$ field as goldstone boson associated with $\widehat{K}_3$ as follows*. First let investigate the initial governing group symmetry and its breaking. We change the obvious $G = O(2)$ symmetry group of circular Kepler orbit with an equivalent form of semi direct product $Z_2 \ltimes SO(2)$.
The main reason for this choice can be traced out in some work on fluid mechanics [23,24]. Although the resulting symmetry group after breaking will be $Z_2$, however there are infinite $Z_2$ symmetries in the plane of rotation respect to the chosen axis. Therefore we add an axis index to $Z_2$ and $SO(2)$ symmetry in semi-direct product:

$$G = [Z_2]_{X_3} \ltimes [SO(2)]_{X_3}$$

Where the indices represent the axis of rotation and $\ltimes$ denotes the semi direct group product.
After breaking the symmetry the original group breaks to $Z_2$, Resulting quotient group obviously will read as:

$$\widehat{G} = G/Z_2 \sim SO(2) \qquad (1.23)$$

Respect to Goldstone theorem these bosons live in the generator space of quotient algebra,

In this case:
$$so(2) \sim u(1) \sim \widehat{K}_3 \qquad (1.24)$$

Therefore $\widehat{K}_3$ presents related Goldstone boson in ground state whose energy is proportional to $\omega_1$ as zero mode. Regarding this breaking symmetry we will deduce the following theorem for space-time broken symmetry. Excited states of Goldstone bosons corresponds the other modes. Consequently the energy states and angular momentum (as the associated charge of $\widehat{K}_3$) are quantized. If $\omega_1$ considered as zero mode, then the energy states and angular momentum eigenvalues can be read as:

$$E_N = N\hbar\omega_1 \, , \quad L = N\hbar \, .$$

Each energy state correspond a geodesic on 2-torus and determined by the general approach in sec 2.2.

### 1.7 Conjectured Theorem:

*If Ensemble of a dynamical setting defined on a topological structure (manifold) embedded in n-dimensional Euclidean space $R^n$ is invariant under group $G$, then boost action on $R^n$ (boost in plane of extra and regular dimension in $R^{n+1}$) breaks the symmetry group $G$ to a subgroup H while the related Hamiltonian and the symmetry group $G \subset \widehat{G}$ remains invariant in $R^{n+1}$, H subgroup is the symmetry of the projection of that manifold orbit onto the R. The G orbit tangent vector(i.e. Killing vectors can be retrieved by a linear combination of Cartan sub algebra of G which recognized as modified basis of a complete symmetry $\widehat{G}$ in $R^{n+1}$.*

Proof of this theorem in the case of $G = SO(2)$ or $SO(3)$ and $\widehat{G} = SO(4)$ is trivial. This theorem can be verified in Lorentz transformation on the Minkowski space-time $M^4$. I propose an inductive Reasoning for general form. Here symmetry breaking takes place in $R^n$ with symmetry group $G$ while the change in $R^{3+1}$ Interpreted as a $\delta\mathfrak{T}$ spanned by the generators specified in $\widehat{G}$ which is a subset of coset generators of quotient group $G/H$.

### 1.8 Chemical potential

Recent method to describe broken generators in the cases with finite charge density will follow the *chemical* potential and effective Hamiltonian. [7,8]. For a free motion on the 2-torus manifold, Hamiltonian could be demonstrated as:

$$\overline{H} = \omega_1 J_1 + \omega_2 J_2 + V = \omega_1 \widehat{K}_3 + \omega_2 \hat{J}_3 + V \qquad (1.25)$$

The basis spanned in the Cartan sub algebra of $so(4)$, moreover should be regarded as the full Hamiltonian realized in $\mathfrak{S}$ equipped with spatial extra dimension. Especially we find out that the orbit space of subgroup expanded by this Cartan subalgebra is a 2-torus with $\widehat{K}_3$ and $\hat{J}_3$ as Killing vectors as basis on its tangential space. By prior assumption in the Kepler problem with $Z_2$ symmetry and 1:1 orbit on 2-torus $\omega_1 = \omega_2 = \omega$, thus $H$ reduces to

$$\overline{H} = \omega(\widehat{K}_3 + \hat{J}_3) + V \qquad (1.26)$$

Without loss of generality by assumption $\omega = \frac{1}{2}$ we get:

$$\overline{H} = \frac{1}{2}(\widehat{K}_3 + \hat{J}_3) + V \qquad (1.27)$$

The vector $\frac{1}{2}(\widehat{K}_3 + \hat{J}_3)$ on tangent space of group manifold (2-torus) represents the direction of the new geodesic trajectory at the origin on 2-torus which also denotes the 1:1 orbit which replaces the original circular orbit after $\delta\mathfrak{T}$ transformation in $\mathfrak{S}$ (the general form will be discussed in the section (2.2).

Furthermore following the method of chemical potential in the cases with finite charge density [5,7,8] the relation of full Hamiltonian $\overline{H}$ after symmetry breaking and $H$ (Hamiltonian in circular orbit) can be introduced as:

$$\overline{H} = H - \mu Q \qquad H = \omega_2 J_3 + V \qquad (1.28)$$

Where $\mu$ denotes the chemical potential and $Q$ stated as some broken generators of the symmetry group $G$. The initial Hamiltonian group symmetry while orbit restricted to two dimension, is $O(2)$ and after a small perturbation on system and breaking this symmetry to $Z_2$ due to the (1.23), (1.24) we will obtain:

$$Q = \widehat{K}_3 \quad , \quad -\mu = \omega_1 = \frac{1}{2}$$

Both Hamiltonian in $C.O.M$ reference frame i.e. ground state reads as:

$$\overline{H} = \frac{1}{2}\widehat{K}_3 + V \ , \quad H = V \qquad (1.29)$$

Equations in Lagrangian formulation is equivalent to [5]:

$$H = \partial_0 \quad , \quad \overline{H} = \partial_0 + i\mu \widehat{K}_3 \qquad (1.30)$$

This is true because in the rest frame the point mass on world line traverse along the time axis with $\partial_0$ generator and after perturbation and jumping to the orbit 1:1 on the 2-torus replaces by $\partial_0 + i\mu\widehat{K}_3$. The situation illustrated in Fig (3b). The concept of jumping from non-degenerate ground state to a degenerate one i.e. "Mexican Cap" implication can be inspired by this configuration. Hamiltonian $H$ is valid in both spaces $M^4$ and $\mathfrak{S}$ but after the action of $\delta\mathfrak{T}$ the valid Hamiltonian changes to $\overline{H}$. Similar method worked-out in recent papers [7, 8] where $\mu$ and $Q$ stand for chemical potential and generator related to NGB respectively. This equations clarifies $\omega_1$ and $\widehat{K}_3$ roles in the model. For an infinitesimal transformation of circular to ellipse orbit (with small eccentricity and $LRL$ vector) the required perturbation has a perturbed Hamiltonian equals to $\omega_1\widehat{K}_3$. The term $\omega_1\widehat{K}_3$ implies the NGB state (Particle). For related *order parameter* the accepted expression as a time-dependent solution reads as [5]:

$$\langle\Phi\rangle(t) = e^{i\mu Qt}\langle\Phi\rangle_0$$

By identification of $\omega_1$ as chemical potential $\mu \sim \omega_1$ and $\widehat{K}_3$ as Goldstone generator $\widehat{K}_3 \sim Q$, by $\nu = \omega_1 t$

$$\langle\Phi\rangle(\nu) = e^{i\omega_1 \widehat{K}_3 t}\langle\Phi\rangle_0 = e^{i\nu \widehat{K}_3}\langle\Phi\rangle_0$$

Reparametrization of time by the perturbation parameter $\varepsilon$ (eccentricity) in diffeomorphism of circular to Elliptic orbit via the linear relation $\varepsilon \propto t$ we can rewritten above equation as:

$$\langle\Phi\rangle(\varepsilon) \cong e^{i\mu Q\varepsilon}\langle\Phi\rangle_0$$

Expansion about small $\varepsilon$: $\qquad\qquad\qquad \delta\langle\Phi\rangle(\varepsilon) \cong (i\mu Q\varepsilon)\langle\Phi\rangle_0$

Comparing this equation with (1.6) clarifies the similarities:

$$|\delta\vec{A}| \propto \varepsilon_3 \widehat{K}_3 \Rightarrow \widehat{K}_3 \sim Q, \varepsilon_3 \sim \mu\varepsilon$$

Setting $\langle\Phi\rangle_0 \sim \vec{K}$ and recalling $\vec{K}$ and $\delta\vec{A}$ from equation (1.5) and (1.6) obtains:

$$\delta\langle\Phi\rangle(\varepsilon) \propto |\delta\vec{A}| \propto (i\mu Q\varepsilon) \qquad (1.31)$$

This reveals the *order parameter* role of *LRL* vector in aforementioned symmetry breaking. The logic of this relation can be understood of the linear relation between the extent of symmetry breaking, order parameter and eccentricity as well as infinitesimal *LRL* vector:

$$LRL \propto \varepsilon \propto \delta\langle\Phi\rangle(\varepsilon)$$

Section.2

2.1 Creation and annihilation operators, string theory

In this model, for a rest reference frame of a system that follows the world line of system (or the center mass of complex system) the dynamic of system restricts to a periodic perpetual motion on a circular geodesic on a cross section of 2-torus surface for 2-outcome system.

Results of this assumption will be coincided, at least partly, with kinematic of string theory as we show it in this section. We construct a complex manifold with three real dimensions for conventional space and one imaginary dimension for extra compacted dimension. For the sake of simplicity the two dimensional complex plane is considered with real axis on the real 3 dimensional manifold and imaginary axis as extra dimension emerges orthogonal to this real manifold. In this scheme the path of particle (or center of mass of system) reduces to combinations of circular curves on this plane.

If the position of center of mass considered as:

$$x^\mu + \frac{p^\mu}{\tau} = x^\mu + \alpha_0 p^\mu \tau \qquad (2.1)$$

Then for the contributions from terms of circular motions in extended space we should sum all terms of complex terms such as:

$$\sum_n r_n e^{i\omega_n \tau} \qquad (2.2)$$

Where $r_n$ denoted as constant complex numbers that determine the initial phases and also represent the "radius" of complex torus constructed of "$n$" angular velocities.

By substitution $z_n = r_n \omega_n$ we obtain final position as a sum of complex terms:

$$\sum_n \frac{z_n}{\omega_n} e^{i\omega_n \tau} \qquad (2.3)$$

So for a general form of coordinates of system position:

$$X^\mu = x^\mu + \alpha_0 p^\mu \tau + \sum_n \frac{z_n}{\omega_n} e^{i\omega_n \tau} \qquad (2.4)$$

This equation looks like the original Taylor expansion adopted in *string theory* with $\frac{z_n}{\omega_n} = \alpha_n$ as modes and $\omega_n$ as "$n$" in Taylor expansion. Obviously the dynamics as well as action resembles the string theory although the main difference will be the point like dynamic in our model and one dimensional string in the other. On the other hand the real projection of oscillator terms onto the real space determines the additional terms of $\Delta x^\mu$ and the absolute value of the imaginary part will be the projection of linear momentum on the real space as additional terms $\Delta p^\mu$. the proof of this statement is simple and omitted in this article.Therefore the simplified form of oscillating terms will be reduced to:

$$\Delta x_n^\mu + i\Delta p_n^\mu = \hat{x}_n^\mu + i\hat{p}_n^\mu \qquad (2.5)$$

Where $\hat{x}_n^\mu$ and $\hat{p}_n^\mu$ denotes the operators of position and momentum and hence $a_n = \hat{x}_n^\mu + i\hat{p}_n^\mu$ and $a_n^\dagger = \hat{x}_n^\mu - i\hat{p}_n^\mu$ Will satisfy the commutation relations of annihilator and creator operators as the quantization process seen in the harmonic oscillators and furthermore in commutation relations in string theory.

2.2 Torsion and contortion as new degree of freedom in extra dimension

It is well known from the original works [25] and [22] that the contorsion $K_{ij}^k$ in Riemann-Cartan space compared to the usual Riemann space acts as a new degree of freedom and couples to spin just like coupling of metric tensor to energy momentum tensor in general relativity.

In usual Einstein-Cartan theory there is two types of curves identified by auto parallel and geodesic respectively derives from the equations:

$$\frac{d^2 x^k}{ds^2} = \begin{Bmatrix} k \\ ij \end{Bmatrix} \frac{dx^i}{ds} \frac{dx^j}{ds}$$

$$\frac{d^2 x^k}{ds^2} = \Gamma_{ij}^k \frac{dx^i}{ds} \frac{dx^j}{ds} \qquad (2.6)$$

In this formalism if the affine connection $\Gamma_{ij}^k$ is symmetric respect to lower indices i.e. $\Gamma_{ij}^k - \Gamma_{ji}^k = 0$ then the torsion and contortion will be vanished and the above two equations coincide:

$$T_{ij}^k = \Gamma_{ij}^k - \Gamma_{ji}^k = 0$$

$$K_{ij}^k = -T_{ij}{}^k + T_{ji}^k - T^k{}_{ij} \qquad (2.7)$$

Now we can imagine a local Minkowski frame for which the Christoffel symbols vanish then:

$$\begin{Bmatrix} k \\ ij \end{Bmatrix} = 0$$

$$\Gamma_{ij}^k = -K_{ij}^k \qquad (2.8)$$

Hence for *parallel transport* of an arbitrary vector $v^i$ in Riemann-Cartan space we obtain [22,26]:

$$dv^k = -\Gamma_{ij}^k v^j dx^i = K_{ij}^k v^j dx^i \qquad (2.9)$$

Without loss of generality and for the sake of simplicity we choose a 2 dimensional spatial sub-manifold of $M^4$; the **xy** plane with rotation group $SO(3)$ and $\hat{J}_3$ as the unique generator of rotation on this plane and an extra dimension imagined to be as perpendicular dimension to this plane with an infinitesimal range on both side of the $xy$ plane.

The emerging from **xy** to extra dimension will be possible through the geodesic on the infinitesimal torus and any particle or center of gravity of any system follows the trajectories on this 2 dimensional torus.

The metric on **xy** plane will be $g_{ij}$ with $\begin{Bmatrix} k \\ ij \end{Bmatrix}$ as connection then the total affine connection $\Gamma_{ij}^k$ will read as:

$$\Gamma_{ij}^k = \begin{Bmatrix} k \\ ij \end{Bmatrix} - K_{ij}^k \qquad (2.10)$$

But in a local Minkowski frame in the vicinity of the point $P$ we have $\begin{Bmatrix} k \\ ij \end{Bmatrix} = 0$ and then again:

$$\Gamma_{ij}^k = -K_{ij}^k \qquad (2.11)$$

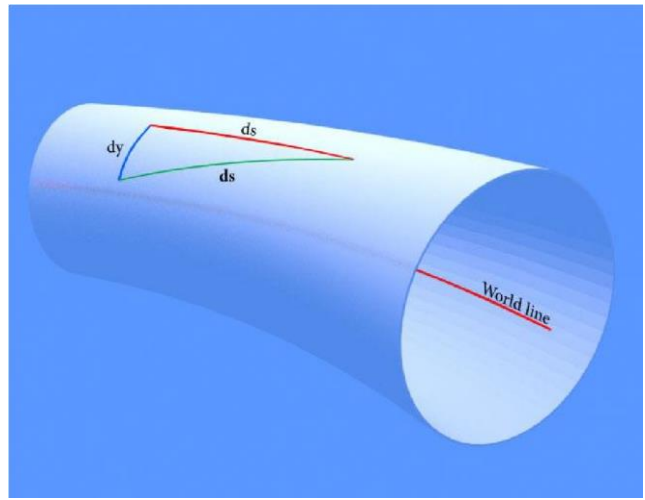

Fig (4): line element on the 2-tori manifold with axis as world line. The green line stated as line element in the 3+1 spatial space.

We verify this relation for an infinitesimal $ds^2$ on the total space i.e. the 2 torus surface Fig (4):

$$ds^2 = g_{\mu\nu}dx^\mu dx^\nu + \varepsilon^2 dv^2 \qquad (2.12)$$

With the definition of $x^0, x^1, x^2, x^3$ the usual space-time coordinate and $x^4 = \varepsilon v$ as the extra dimension which restricted to small amounts on infinitesimal 2 torus with a fixed small radius $\varepsilon$ and $v$ as the phase parameter. In the rest frame the rotation of particle assumes generated by $\widehat{K}_3$. For a geodesic of 1:1 winding number the coefficient of generators $\hat{J}_3$ and $\widehat{K}_3$ should supposedly be equal, then the appropriate algebraic base in $so(4)$ will be: $\qquad A_i = \frac{1}{2}J_i + \frac{1}{2}K_i \qquad , B_i = \frac{1}{2}J_i - \frac{1}{2}K_i$

This conveys us that the equal coefficient of generators in changing base from original $so(4)$ to modified one is not accidental. This modified base also is well known method of decomposition of $so(4)$ algebra in Hydrogen atom problem.

Return back to the equations of auto-parallel and geodesic equation by the:

$$\frac{d^2 x^k}{ds^2} = \Gamma^k_{ij}\frac{dx^i}{ds}\frac{dx^j}{ds} - K^k_{pq}\frac{dx^p}{ds}\frac{dx^q}{ds} \qquad (2.13)$$

$\begin{Bmatrix} k \\ ij \end{Bmatrix}$ Indices and $p, q$ takes the values 1, 2, and 4. For compatibility of indices with the extra dimension we simply assume $x^0 = t$ as a parameter and $v = \omega_1 t$:

$$x^4 = \varepsilon \omega_1 t = \varepsilon \omega_1 x^0 \qquad (2.14)$$

So the extra dimension as a new degree of freedom reduces to a linear function of $x^0$. Therefore the extra dimension will absorb to $x^0 = t$ in the equation

$$ds^2 = g_{\mu\nu}dx^\mu dx^\nu + \varepsilon^2 \omega_1^2 (dx^0)^2 \qquad (2.15)$$

For Kepler problem we use the Schwarzschild metric for a gravity field with spherical symmetry:

$$ds^2 = -\left(1-\frac{r_s}{r}\right)c^2 dt^2 + \left(1-\frac{r_s}{r}\right)^{-1} dr^2 + r^2(d\theta^2 + \sin^2\theta d\varphi^2) \qquad (2.16)$$

The resulting differential element of extra dimension respect to above assumptions will be: $dy = \varepsilon dv$ hence the metric element for a spherical symmetric field with source field on the center of coordinate and on the surface $\theta = \frac{\pi}{2}$ in spherical coordinate will be read as [16, 17]:

$$ds^2 = -\left(1-\frac{r_s}{r}\right)c^2 dt^2 + \left(1-\frac{r_s}{r}\right)^{-1} dr^2 + r^2 d\varphi^2 + dy^2 \qquad (2.17)$$

Substituting $= \varepsilon dv$ :

$$ds^2 = -\left(1-\frac{r_s}{r}\right)c^2 dt^2 + \left(1-\frac{r_s}{r}\right)^{-1} dr^2 + r^2 d\varphi^2 + \varepsilon^2 dv^2 \qquad (2.18)$$

By introduction of $\omega_1$ as angular velocity along extra dimension we have:

$$ds^2 = -\left(1-\frac{r_s}{r}\right)c^2 dt^2 + \left(1-\frac{r_s}{r}\right)^{-1} dr^2 + r^2 d\varphi^2 + \varepsilon^2 \omega_1^2 dt^2 \qquad (2.19)$$

Substituting $\frac{r_s}{r} = \frac{2\omega_2^2 r^2}{c^2} = \frac{2m}{r}$ and $\omega_1 = \omega_2 = \omega$

$$ds^2 = -\left(1 - \frac{2\omega^2 r^2}{c^2} - \frac{\varepsilon^2 \omega^2}{c^2}\right)c^2 dt^2 + \left(1-\frac{r_s}{r}\right)^{-1} dr^2 + r^2 d\varphi^2 \qquad (2.20)$$

It is clear that the added component is small compared with $\frac{2\omega_2^2 r^2}{c^2}$ since:

$$\varepsilon \ll r$$

The contribution term of extra dimension $\varepsilon^2 \omega_1^2$ in above equation, is confined to $x^0 = t$. This confinement also can be observed through the conditions in Kaluza Klein theory with torsion in extra dimension and related papers [25, ] as:

$$\mathbf{g}_{\mu\nu} = g_{\mu\nu} + eA_\mu A_\nu \Phi^2 \qquad e = \pm 1 \qquad (2.21)$$

Where $A_\mu = 0$ for $\mu \neq 4$ hence the contributed term will reduce to $(A_4)^2 \Phi^2$ which is comparable to $\varepsilon^2 \omega_1^2$. Here an interesting realization can be shown by introducing $A_4$ as Higgs like boson that corresponds to radial degree of freedom $\varepsilon$ and $\Phi$ scalar potential as Goldstone boson momenta equivalent to $\omega_1$. This interpretation coincides the result of Lagrangian in (1.20) and field $\nu$ as Goldstone boson.

Comparing a generalized geodesic equation in Riemman-Cartan space time with Weitzenboch connection:

$$\Gamma_{ij}^s = \begin{Bmatrix} s \\ ij \end{Bmatrix} - K_{ij}^s \qquad (2.22)$$

This equation reveals that the additional term in coefficient $c^2 dt^2$ plays a role similar to *contorsion* $K_{ij}^s$ in geodesics equation with torsion [27, 25].

Involving contorsion in the model we need first to calculate the three index Christoffel symbol with the above mentioned correction:

$$\begin{Bmatrix} s \\ ij \end{Bmatrix} = \frac{1}{2} g^{sk} \left( \frac{\partial g_{jk}}{\partial x^i} + \frac{\partial g_{ki}}{\partial x^j} - \frac{\partial g_{ij}}{\partial x^k} \right) \qquad (2.23)$$

It can be shown that this correction only effects on $\begin{Bmatrix} 0 \\ 10 \end{Bmatrix}$ since this term directly depends on the coefficient of $c^2 dt^2$ in metric equation and its derivative respect to $r$ [18]. If we assign this coefficient as $\beta$ then in usual space-time coordinate [4]:

$$\begin{Bmatrix} 0 \\ 10 \end{Bmatrix} = \frac{\beta}{2\beta'} = \frac{m}{r(r-2m)} \qquad (2.24)$$

Now with the additional term of $\beta$ in equation (2.20) it will be read:

$$\Gamma_{10}^0 - \begin{Bmatrix} 0 \\ 10 \end{Bmatrix} = K_{10}^0 = -\frac{\varepsilon^2}{8r} \qquad (2.25)$$

$$\frac{d^2 t}{ds^2} + \left[ \frac{2m}{r(r-2m)} - K_{10}^0 \right] \frac{dr}{ds} \frac{dt}{ds} = 0$$

Therefore the unique geodesic equation that changes by the additional term reads as:

$$\frac{d^2 t}{ds^2} + \left[ \frac{2m}{r(r-2m)} + \frac{\varepsilon^2}{8r} \right] \frac{dr}{ds} \frac{dt}{ds} = 0 \qquad (2.26)$$

Then the geometry of the model requires a limited contorsion $K_{ij}^s$ with vanished components except the corresponding component to proper time i.e. in this setting the contorsion $K_{10}^0$ interpreted as the term in (2.25) realized in extra dimension.

## 3. Conclusion

We introduced a new interpretation of *LRL* vector as order parameter through a type of symmetry breaking of $O(2)$ to $Z_2$ in configuration of an orbiting point mass in a spherical gravitational field.
The breaking symmetry deduced on the basis of 'Ensemble view' of infinite copies of the periodic system in degenerate states. Connection between this breaking symmetry and limited boost like transformation induced by generators of $so(4)$ algebra and related parallel displacement and covariant derivative has been explored which leads to exploit the Cartan torsion and contorsion in the model. We introduce a "generalized angular momentum" and use a new degree of freedom explained in extra dimension on 2-torus as a perpetual rotation which is comparable to the recently proposed perpetual motion of Wilczek.et.al and explore the relation of this motion with $so(4)$. Moreover we define the metric on 2-toral hyperspace which leads to new term in Schwarzschild equation with neglecting impression compared to other terms. This approach also reveals a gauge field for gravity in the problem. Symmetry breaking considered as a result of a perturbation realized in hyperspace with perturbing Hamiltonian operator expanded by some broken generators of $so(4)$ governing algebra and order parameters known as chemical potentials. This method also suggest the second type of massive Nambu-Goldstone boson with the energy of perturbing Hamiltonian which mediate the interaction leading to transformation of circular to ellipsoid orbits. So the transition from non- degenerate circular state to degenerate ellipse orbit can be visualized as changing the potential well from normal to Mexican hat. On this background in next work we generalize this method by introducing a modified basis $so(4)$ algebra to approach the dynamical systems modelled by free motion on negative curvature as well as hyperbolic modular spaces for exploring the connections of Golden ratio, Riemann hypothesis and Zitterbewegung with this model. We will show in the next paper that the charges in root space representation of modified $so(4)$ determine the physical probability of system through an exponential map on the group manifold.

Acknowledgments

I would like to thank Dr. N.Nafari, Dr.S.Rahvar and Dr.M.R.Rahimitabar for their precious comments and guidelines.